# Users' participation to the design process in an Open Source Software online community

Flore Barcellini [(1)], Françoise Détienne [(1)], Jean-Marie Burkhardt [(1,2)]

(1)INRIA Eiffel Group, Rocquencourt, France
(2)Université Paris 5, Laboratoire d'Ergonomie Informatique, Paris France
Flore.Barcellini, Francoise.Detienne ,Jean-Marie.Burkhardt@inria.fr

**Abstract.** The objective of this research is to analyse the ways members of open-source software communities participate in design. In particular we focus on how users of an Open Source (OS) programming language (Python) participate in adding new functionalities to the language. Indeed, in the OS communities, users are highly skilled in computer sciences; they do not correspond to the common representation of end-users and can potentially participate to the design process. Our study characterizes the Python galaxy and analyses a formal process to introduce new functionalities to the language called *Python Enhancement Proposal* (PEP) from the idea of language evolution to the PEP implementation. The analysis of a particular pushed-by-users PEP from one application domain community (financial), shows: that the design process is distributed and specialized between online and physical interactions spaces; and there are some cross participants between users and developers communities which may reveal boundary spanners roles.

## 1    Introduction

Open Source Software (OSS) is software that can be run, distributed, studied, changed and improved by its users thanks to a specific license (www.gnu.org). The design behind OSS (OSSD) becomes an important phenomenon in the computer science world: there are thousands of OSS projects and millions of users of OSS systems. However, in this world, users do not correspond to the common representation of "unskilled" end-users of computer applications [1]. They can be highly skilled in computer sciences even if they use OSS for different application domains (education, biology, scientific computing...).

OSS can often involve a distant and asynchronous form of computer-supported collaborative design, and a large number of developers, members of online communities [2]. If there is a lot of research on collaborative design (e.g. [3][4]), very few focused on distant and asynchronous design situations. We assume that OSS design process is an interesting paradigmatic case to study distant and asynchronous collaborative work.

Moreover OSSD is a case of continuous design [5][6] and participatory design [7]. In traditional software design models, different phases are distinguished (design, implementation, production). OSSD mixes these different phases and does not elaborate stop criteria to the design process: design becomes continuous, new functionalities can always be proposed and discussed whatever the step in the project. Whereas users, in traditional user-centred design models, take parts in the design process as informants - in the functional analysis phase- or as evaluators - in the prototype and simulation phases- in OSSD, users can be potentially involved in all the phases of the design process. This participation is seen as one of the most important factor explaining the success and the quality of the designed OSS. Thus OSSD can be considered as a participatory kind of design.

Due to these two last characteristics, continuous and participatory design, OSS online communities are of particular interest. Forms of participation in these communities are supposed to be « open » in time (the design is continuous) and for different kind of participants whatever they are (administrators, developers, or users). As far as we know, there is no research that aims at understanding globally the OSSD process and the position effectively occupied by users proposing new functionalities in this self-organised design process.

In this direction, the objective of this research is to understand the OSS design process by analysing the ways members of OSS communities, and especially users, participate in this process. This research is focused on a major OSS project called Python, which is a programming language.

In the following sections, after a state of the art of prior studies highlighting OSSD main characteristics, we will present our research questions and strategy. Then, we will present the results of this research being still in progress. Finally, we will present the perspectives of this research.

## 2  OSSD main characteristics

OSSD is a mediated, distant, asynchronous and emerging collaborative design process, for which a few general characteristics have been highlighted in recent studies [8][5][9][10], especially for major projects.

**A distributed and emerging design process.** The OSSD process is distributed among three spaces on the Internet [10][11]: a discussion space (mailing lists, forums, chat…), a documentation space (project web site, related websites, blog, wiki, online documentation…), and an implementation space (source code and its versions[1]). Internet tools such as mailing lists, forums, wikis, blogs, documents and code versions repository (CVS or Subversion) and GroupWare mediate the major part of the design process.

In theory, OSSD has an emerging design process, which is not formally defined a priori. In the case of "small" projects (i.e. developers < 10 persons), work is not assigned and people undertake the work they choose to. Bigger projects seem to proceed differently: work is assigned and some formal and explicit processes of coordination may be elaborated from practices [9].

---

[1] Supported by the Concurrent Version System (CVS) or by Subversion

In a previous work [11], we have studied such a formal process in the Python community, which is one of the major OSS project. This process, called Python Enhancement Proposal (PEP) is used to propose language evolutions. It is similar to methodologies followed in traditional software design projects, such as *Technical Review Meeting* or *Request for Comment* [12]. This study focuses mainly on the discussion space and on the python-dev mailing-list in which the major part of the PEP design activity is expected to occur. Parts of the design choices and alternatives, together with their rationales are elaborated and discussed in this space.

We found that online design discussions are focused and framed by specific members of the project, especially the project leader; and that the proportion of design activities are different in online interactions compared to face-to-face interactions: for instance clarification activity occurs as in face-to-face design situations [13] but seems to be framed by the project leader and reserved to specific locations in the discussion space.

**OSS communities and participants.** According to OSS movement leaders (e.g. [14]), as well as some scientific studies, OSSD projects are structured as communities (e.g. [15]). More specifically, OSS project can be seen as online communities according to Preece's definition [1]: on one hand, their members constitute a group of people connecting together on the Internet with a common goal- to develop a software; on the other hand, this group is led by the OSS ideology norms and principles (e.g. [16]). For instance, OSS designers believe that the OSS way of design is more effective than the proprietary software approach.

OSS projects can also be seen as epistemic communities [15]. Indeed, OSS members have the "meta"- objective of producing and constructing knowledge about the artefact they develop, for the benefit of all the community. In that sense, their objective is not only to gain individual knowledge but also to co-construct and to share knowledge. Finally, OSS projects are also meritocratic communities, sometimes highly hierarchical, at least in the case of major OSS projects [17][18].

Major OSS projects are constituted by a huge number of volunteers, whose average profile is as follows: mainly male (98%), under 30 years old, living in Northern America and Europe, and with generally a high level of qualification in computer sciences or information technologies [19].

Five different statuses are currently distinguished in OSS projects. The *project leader* - sometimes referred to (semi)-ironically as the BDFL (Benevolent Dictator For Life)- is generally the creator of the project as in Python. The *core team* or *administrators* have to maintain the code base, the documentation, and they can often make executive decision. The *developers* participate to the evolution of the OSS and maintain some of its parts, they have the rights to change the source code. *Active users* participate in mailing-lists discussions as informants for newcomers for instance, report bugs and/or correct these bugs with patches; in this case they are *patchers* or *contributors;* and propose new modules. These users are considered as the main force of OSS designs process versus proprietary one: most bugs are detected and fixed because "they are many eyeballs looking at the problem" [14]. Other users are called *passive users* as they only use the software or lurk on the Internet [20].

## 3   Research Issue

**Users in OSS design.** Commonly in software design, users can be involved in 3 levels of the design process: the *needs and requirements elicitation* using human-centred design methodologies such as field studies and interviews, or participatory design; the proper *design* (generation and choice of design solutions) as in participatory design [7]; and in *evaluation* usability tests for instance.

In the case of OSS, these 3 levels seem to be integrated, as active users, developers and administrators are *all users* and can be *designers* of the software. Users can modify the code, thus they can satisfy their own needs and requirements proposing new modules, at least if they have the skills to do so. They are free to participate in discussions and evolutions of the language at least if they are able to be integrated in the project. Indeed, several studies (e.g. [18]) point out that, to be integrated into an OSS projects, a participant has to proof his value to his peers by his appearing technical skills and ability to engage and maintain online discussions. Finally, they can report bugs and correct them so the software is continuously evaluated and corrected, contributing to increase a certain aspect of the software usability. We have to precise that this model leads to consider a certain aspect of usablility, i.e. the one of users that have the skills to report or correct bugs, but may excluded *unskilled end-users*. As pointed by Twidale and Nichols [21] the OSS design process is not yet well armed to integrated software usability for end-users.

**Research questions.** They have been several research on OSS users participation in bug reporting (e.g. [22]), but as far as we know none as focus on the OSS users participation to the proper design process, i.e. proposing new functionalities for the software and generating and choosing design solutions. Beside the idealistic picture that users may intervene freely in the process, we will question whether users who are neither administrators nor developers in the core Python community can really have an impact in the design choices and decisions. In this direction and from a cognitive ergonomics point of view, our research questions are as follow:

- *In what parts of an OSS project are the various participants (administrators, developers, active users) involved?*
- *In which spaces (mailing-list or others spaces) do users elicitate their needs and requirements for new functionalities?*
- *What are the users forms of participation during these elicitations?*
- *What are the users forms of participations during the design process of these new functionalities? When does this occur in the process?*
- *Are they key participants that are linking users and administrators-developers communities? When does this occur in the process?*

The links evoked in this last issue can be realised by some specific participants that may act as "boundary spanner" for instance [23]. We assume that *cross-participation* can reveal these forms of links between the two mailing-lists and thus between users, administrators and developers. We define *cross-participation* as participation at same-topic discussions, occurring in parallel in mailing-lists. This notion is linked to *cross-posting* defined as "the practice of posting the same message to multiple newsgroups" [24] and as "broadcast interactions to multiple newsgroups" [25]. We extend this to cross-participation considering that these interactions can be more than the post of the same message but also a reformulation of the main ideas content in a message.

**Research strategy.** In this perspective, we need to have a global view of what is happening from the idea of evolution proposed by users and/or developers and emerging from the discussion space or other interaction spaces, through its acceptance within the community, its specification, and finally its implementation. Thus, we choose to go on focusing our research on the Python community and its PEP process (use to propose language evolution) as a relevant entry to trace the design process in this community.

A part of this study will be focused on the discussion space in which the major part of the design occurs [9]. We will analyse both the python-dev and the python-list mailing-lists. The python-dev mailing-list has already been studied in our previous paper [11] and is focused on design and maintenance issues. Indeed participation of users is expected in the python-list mailing-list, which deals with general discussions on Python.

## 4     Study of a pushed-by- users design process

In this section, we will first present the PEP process on which we focus on and the method we develop to address our research questions. Then, we will present our first results in this direction.

### 4.1    Focus on the PEP process

The *Python Enhancement Proposal* (PEP) process is an explicit means for proposing new language evolutions, for collecting community input on an issue, and for documenting chosen design decisions. A PEP is a document written to describe a new language feature, and the associated discussion process. The PEP design process is constituted by a set of activities that take place in the three different spaces [11]: the discussion space, the documentation space, and the implementation space. Once *PEPs editors* (one administrator and one developer) have accepted a rough-draft PEP, the author of the PEP, called the *champion*, is responsible for posting the PEP to the community forums and mailing lists where the PEP is discussed. Archives of discussion, decisions regarding the PEP, and the different versions of PEP are kept in the documentation space. Information about the PEP and its status is, therefore, distributed between these two spaces. After a PEP has been accepted, it is given a final review by the leader of the Python project. Finally, a new piece of code is written to implement the PEP. This code is integrated into the project's code archive: the implementation space.

### 4.2 Method

To fulfill our research questions we first need to: *(1)* have an overview of the Python project, its application domains and the participation of users to them; *(2)* identify and select some idea of evolution that would evolve to a "pushed-by-users" PEP; *(3)* obtain and analyze traces of exchanges around these proposals in the discussion space. To address our first two issues we conducted some field interviews with users of Python in France. To address the last point, we analyse online exchanges, on a particular evolution idea and its related PEP, between participants in the Python community in the two main mailing-lists of the project: *python-dev* and *python-list*.

**Field interviews.** We have interviewed ten active users of the Python language in France. These people were selected according to two criteria: their involvement into the French and European Python community and / or their participation to the French and the general python mailing-lists. Six from these ten persons are working in different application domains: 2 in computer science institutions, 2 in biology related institutions, 2 in nuclear institutions. The four remaining are working in little firms providing services around Python[2].

These semi-structured interviews addressed different issues: the place and advantages of Python in the computer science world; the participation of users in the community; the contribution of users to Python; the resources that help users staying aware of the language evolution and the news regarding the language and the community; and the evolutions of the language (PEPs), which have been most significant for them or for some other users in the community. To help interviewees addressing this latest issue, they were presented with some elements of the documentation space, i.e. a webpage of the Python site summarising the new PEPs. The interviewees were free to comment the PEPs they chose to on this webpage.

Interviews were then transcribed and analysed to highlight the following issues: Python and its related application domain, users' contributions to the project; their modes of interactions. A list of interesting "pushed-by-users' PEPs was also outlined this way. The data collected and this analyse enabled us to obtain schemas formalising the Python galaxy and its interactions spaces that will be presented in the next section.

**Deeper analyses of the PEP design process.** We selected one of the PEP (PEP 327) proposed by users as significant for them or other users during the interviews. This PEP is related to money and decimal issues in the Python language.

To study this PEP process, we coupled: (1) an analysis of the design' traces available online - the discussions in the python-list and the python-dev mailing-lists and the weblog of the champion of the PEP- with (2) an interview with the champion of this PEP. In the following, the champion will be called user-champion to remind that is a user before being the champion of this PEP.

---

[2] As far as we know there is no study characterising Python users in France and around the world, we are thus not able to say if this sample is representative of this population. But at this step of our research our objective is more to go further in the understanding of their participation than to generalize our results.

As all discussions are publicly available and archived online, the data was gathered by searching from the python-list and the python-dev mailing-lists for the keywords: *decimal*, *money*, *currency*, *PEP 327* and the name of the user-champion. The search was performed from the first post of the user champion in October 2003 to May 2006. As O'Shea and Exton assume [26] it was not possible to automate the data gathering process as each message which was returned by the keyword required reading by the first author of this paper to ensure it was indeed a message dealing with the design issue we are interested in.

The PEP 327 corpus is composed of 52 discussions in the python-list and the python-dev from the 17th of October 2003 to the 23th of May 2006. In the weblog, we found 5 articles referencing to Python conferences' sprints related to the design process we are interested in.

To characterize the temporal organization of the process, we identified: *(1)* the date and time at which they occurred; *(2)* the 5 design issues or themes (T) addressed by discussions, thanks to the subject of the message header and the first reading of its content. Theme 1 and 2 (T1 and T2) are about money issues related to the PEP, T3 is about the pre-PEP proposal on money and decimal data type, T4 deals with the discussions about the accepted PEP proposal, and T5 is composed of discussions of different issues on decimal.

To characterize the users forms of participations, we identified: *(1)* the participants and their status (project leader; administrators, developers, the user-champion and other users). We call *users* those that are not clearly identified as *administrators* or *developers* on the project webpage[3]; *(2)* their participation according to the number of messages posted in all discussions. Thus, we distinguish between High and Low participants according to the median number of messages posted (median=1) (High Participant (HP) sent 2 or more messages, and Low Participants (LP) posted one message); *(3)* the degree of involvement of participants in all the process, i.e. the number of discussions in which they are present; *(4)* the presence of *cross-participants* in the same discussions of the two mailing-lists.

### 4.3 Results

In this section, we present the results obtained so far in this current research. The first results deal with the understanding of both: the "Python Galaxy" - the Python programming language and its different application domains, but also the interaction spaces in which activities of the community is distributed. The second part of this work is a more fine-grain analysis which aim at characterizing the extended PEP process related to a specific application domain.

**The python galaxy**. Interviews and analyses of some pieces of the documentation space of Python led us to better understand the "Python Galaxy" in which users can participate (figure 1).

---

[3] There is actually a list of administrators and official developers.

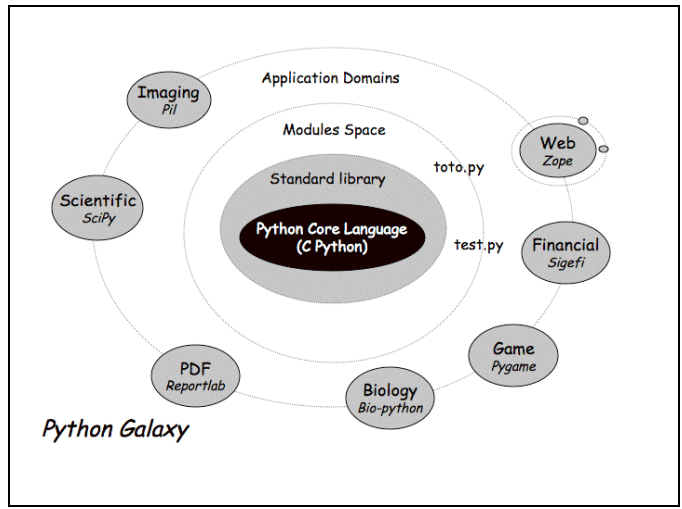

**Figure 1**. The *Python galaxy*

The Python programming language itself and its standard library occupy the core of this galaxy. This core is *a priori* the place for Python's core team and developers. But users can also act in this world by participating in online discussions, by reporting bugs but also by "contributing" to the evolution of Python by proposing "patches" (bugs corrections). In theory, they can also ask for language evolutions, using the PEP process for major ones. In this case, the demander has to propose code and to "champion" his/her idea to increase the chance for his/her request of evolution to be considered and to succeed within the community. The core of Python is strongly related to the *implementation of Python* projects (projects that aim at writing Python in Java, C++ or even in Python).

Around this core, active users of the Python programming language are developers of modules based on Python, to satisfy specific needs of different application domains. Some of these modules may become parts of the core language; other can remain isolated ones. They are available on private or companies' websites (such as Activestate Cookbook for instance), or some index such as PyPi (official index system for python modules).

Python users can also participate into project related to various applications domains (web development, scientific computing, biology, and financial…), which proposed collections – packages - of modules dedicated to these applications. They often belong to communities with their own projects websites and groupware and members of these communities may be related to end-users of these applications based on Python.

**Extension of the interaction spaces.** As we pointed out previously, OSSD process is distributed in three spaces on the internet (documentation, implementation and discussion). Members of the Python community can participate in these different spaces. In the discussion space, they have some dedicated mailing-list (python-list for general discussions about Python, python-tutor for newcomers) but also special interest group (sig) mailing-lists such as the education mailing –list for people interested in Python modules dedicated to educational needs, or specific sub-communities mailing lists such as the numerical python projects. The documentation space is very important for participants. There are several websites providing information about Python: the official Python website, companies' websites, blogs and wikis related to python, all the archives on the online discussions... In the implementation space, participants can act in different ways: bug reporting, patching, providing modules. But some users not only interact and act on the language in these online space, but also in face-to-face meetings such as annual international conference (EuroPython, RMLL, PyCon...) or *sprints* which are small meetings organized by project to address some specific programming issues using XP and Agile methodologies (http://www.agilealliance.org/).

Thus, participants to the Python project are not only present in *online interactions spaces* but also in *physical interactions spaces*. The OSS design process is thus distributed between four spaces: 3 online (documentation, discussion and implementation) and one physical. Another issue here is to characterize the relative importance of interactions in these spaces for the design process and between members of the community (project leader, administrators, developers, users from different application domains).

**Before, during and after a "pushed-by-users" PEP.** We focus on the Python programming language world but on a PEP, which have some links with a specific application domain (financial one). We present first a temporal representation of the extended PEP process and then we will characterize who are the participants in this process and in which interactions they participate.

Figure 3 represents a temporal view of the design process. The python-dev and the python-list discussions are represented in parallel. Each discussion and its themes, is represented by a symbol. Conferences in which they have been some interactions around the design process are represented by vertical grey line.

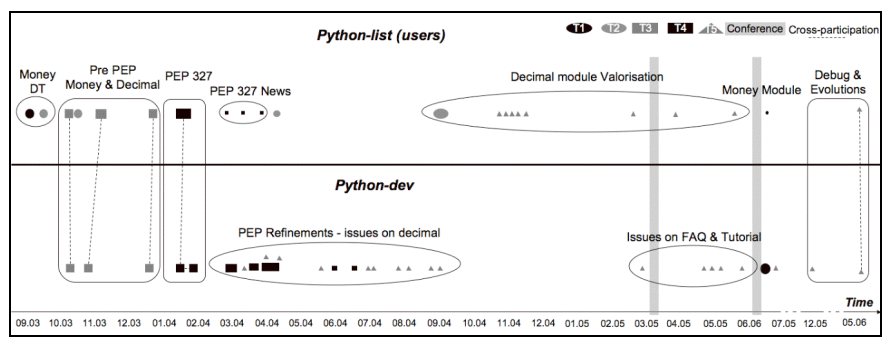

**Figure 2**. Temporal view of the PEP 327 design process

Eight main steps can be distinguished in the figure. (1) *Money DT phase*: The process begun, in October 2003 with a post in the python-list by a user. This post concerned the need for a type suitable for money operations and is linking to financial application domain. This user is indeed the project leader of a financial OSS project. (2) *Pre PEP money and decimal phase*: Following this discussion and the comments he received the user proposed a pre-PEP called *money data type* to the python-list and the python-dev mailing-lists. It appears that to create a money data type requires work on the decimal type of Python first of all and the proposal evolved to a prePEP called *decimal data type*. (3) *PEP 327 phase*: In the next step the pre-PEP was accepted, acquired a status and a number (PEP 327) with the user as a champion (called user-champion in the following). (4) *PEP Refinements-issues on decimal phase*: The new PEP kept on being discussed, such as related issues on the decimal type but only in the python-dev mailing-lists. In parallel, some informative isolated posts appeared on the python-list but they didn't lead to discussions. (5) *Decimal module valorisation*: After this step, different issues and questions emerged about decimals or money on the python-list and the user-champion documented them referring to "his" new decimal type and module (see *PEP valorisation*). This step may be crucial considering the traffic of the mailing-list, the turn-over of participants and the temporality of the process (several month passed between his first post). (6) The following step "*Issues on FAQ and tutorials*" appears only in the python-dev mailing-list and deals with general questions such as the decimal module documentations and FAQs, that may appear during the module valorisation phase within the python-list and be relayed to administrators and developers (7) Finally the champion announced that he implemented a money module (his first need) and that it was publicly available on *sourceforge*. Analysing traces from his weblog outlined that specifications of this module were discussed at the *PyCon* conference in March 2005 and coded in a sprint at *EuroPython* conference in June 2005, as indicated by the vertical grey line on the figure. (8) The ultimate step so far correspond to a *bug correction and evolution* of the PEP, following the evolution of current standard on decimal and some bugs corrections.

This analysis highlights that there was two main steps in this online design process coupled with a face-to-face design step at conferences. In the first step, there were parallel same-topic discussions in the two mailing-lists, with participants that cross-participated to these discussions, as we will see in the next section. The second step occurred as the process advanced, specialized discussions appearing in the two mailing-lists in accordance either with the design perspective on python-dev and the use perspective of python-list (valorisation of the decimal module for instance).

**Participation view.** Table 1 summarizes the number of participants according to their status and their participation in number of messages posted to the discussions of the python-list and the python-dev mailing-lists.

**Table 1.** Participants and participation in number of posted messages to the discussions of the two mailing-lists according to the status

| Status | Level of part. | Distribution of participants | | Distribution of posted messages | |
|--------|----------------|-------------|------------|-------------|------------|
|        |                | Python-list | Python-dev | Python-list | Python-dev |
| Leader | PL             | 0 (0%)      | 1 (2%)     | 0 (0%)      | 19 (5%)    |
| Admin  | HP-Ad          | 1 (1%)      | 4 (8%)     | 9 (3%)      | 86 (21%)   |
|        | LP-Ad          | 1 (1%)      | 0 (0%)     | 1 (0%)      | 0 (0%)     |
| Dev    | HP-Dev         | 2 (4%)      | 12 (25%)   | 51 (15%)    | 93 (23%)   |
|        | LP-Dev         | 4 (4%)      | 6 (13%)    | 4 (1%)      | 6 (1%)     |
| Users  | Champion       | 1 (1%)      | 1 (2%)     | 70 (21%)    | 99 (24%)   |
|        | HP-Us          | 29 (30%)    | 17 (35%)   | 145 (43%)   | 95 (23%)   |
|        | LP-Us          | 58 (60%)    | 7 (15%)    | 58 (17%)    | 7 (2%)     |
|        | *Total*        | *98*        | *48*       | *338*       | *405*      |

Several patterns of participation can be outlined: *(1)* The pattern of users is not the same in the two mailing-lists: in python-list, users are mainly low participants (60%, 58/98) who participate only once in the discussions, whereas in Python-dev users are mainly high participants (35%, 17/48) and produce 23% (95/405) of messages. RD[4] confirms that HP-users tend to participate more in the Python-dev than in the Python-list, and conversely for LP users. *(2)* The user-champion is highly present in the two mailing-list posting 25% (70/338) of messages in the python-list and 24% (99/405) in the python-dev and RD reveals that he tends to participate more on the python-dev mailing-list. This result is consistent with our previous results [10] in which we outlined that the champion was boosting the discussions of python-dev to make the community reach to a consensus. This point extends this result to the python-list. However, he might participate more on the python-dev due both to the relevance of design issues discussed in this list and the fact that he has to proof his value to other developers. *(3)* Developers are more present in the python-dev (38%, 18/48) mailing-list than in the python-list (8%, 6/98), and they tend to participate more in Python-dev as they post 24% (41/155). This is consistent with the python-dev implicit rules of participation, i.e. this list is mainly for administrators and developers who have, in theory, to be invited to participate by the PL. *(4)* An HP administrator is present in the python-list. He is in fact the specialist in scientific computing and champion of the solution used before the PEP for decimal and monetary issues. In the python-dev 4 administrators out of the 6 of the project are present and are all high participants: they post 21%(86/405) of messages *(5)* The project leader does not participate at all to the python-list and participate less (5%, 19/405) in python-dev than in others discussions studied in our previous paper [10], in which he posted as far as 20% of messages. He is indeed relayed by the HP-A specialist of scientific computing. This point confirms that the PL and administrators are complementary in the project.

---
[4] RDs measure the association between two nominal variables. There is attraction when the RD is positive, and repulsion – when it is negative. By convention, we reported only attractions (and sometime repulsion) with values >.20 (<-.20).

To characterize cross-participation, we first identified 13 people who post in the two mailing lists all along the discussions: the user-Champion, two administrators (from which one is a specialist in scientific computing who contributed a lot to this process as declared by the user-champion), five developers (from which one Dev1 is the former champion of a decimal module, one contributed a lot to the process, and one is a PEP editor), and five users (table 2).

**Table 2.** Number of posted messages and presence in discussions for participants present in the python-dev and the python-list.

| Status | Level of part. | Distribution of posted messages | | Presence in discussions | |
|---|---|---|---|---|---|
| | | Python-list | Python-dev | Python-list | Python-dev |
| Ad | Ad1 | 9 | 52 | 5/22(23%) | 14/30(47%) |
| | Ad2 | 1 | 27 | 1/22(5%) | 12/30(40%) |
| Dev | Dev1 | 16 | 28 | 4/22(18%) | 12/30(40%) |
| | Dev2 | 35 | 6 | 5/22(23%) | 9/30(30%) |
| | Dev3[5] | 1 | 4 | 1/22(5%) | 2/30(7%) |
| | Dev4 | 1 | 22 | 1/22(5%) | 9/30(30%) |
| | Dev5 | 5 | 12 | 1/22(5%) | 5/30(17%) |
| User | Champion | 70 | 99 | 21/22(95%) | 28/30(93%) |
| | Us1 | 1 | 1 | 1/22(5%) | 1/30(3%) |
| | Us2 | 2 | 9 | 2/22(9%) | 4/30(13%) |
| | Us3 | 3 | 4 | 1/22(5%) | 1/30(3%) |
| | Us4 | 1 | 2 | 1/22(5%) | 1/30(3%) |
| | Us5 | 1 | 1 | 1/22(5%) | 1/30(3%) |
| | *Total* | *146* | *267* | | |

The user-champion participates to more than 90% all discussions in the Python-dev (28/30 discussions, 93%) and the python-list (21/22 discussion, 95%). Ad1 (5/22), Dev1 and Dev2 (4 and 5/22) are also present in quite 20% of python-list discussions (especially in first discussions), and around 40% of discussions for python-dev (14/30 for Ad1 and 12/30 for Dev1), 30% for Dev2 (9/30). In the python-dev, we notice that Ad2 (one contributor to the PEP) is present in 40% (12/30) of discussions, such as Dev4 (9/30, 30%). Out of the 13 people posting in the two mailing-litsts, these 5 (User-Champion, Dev1, Dev2, Dev4 and Ad1) who are highly present in discussions are also cross-participants to same discussions in the two mailing-lists.

These results confirm that there are some roles specializations into the project (not all the administrators, nor the project leader are cross-participants for instance) and that some "key" participants (users and developers) seem to be more involved. An issue is here to characterize the nature of their involvements and their activities online to precise these forms of participation.

---

[5] Dev3 is one of the PEP editors

**Synthesis.** Figure 3 displays a summary of this process: the need for a *money data type* emerges in the application domain (financial) space and is relayed by a user (the future champion) to the Python language discussion space (python-list) (step 1 on the figure). After several interactions in the python-dev and python-list mailing-lists, the first need is transformed and a more generic PEP (dealing with decimal) is elaborated such as a decimal module (step 2). The initial need for a money data type is still going on and is specified during a conference on Python. At the following conference, the module is primarily coded and the creation of a money module project was announced (number 3). This leads to the creation of a PyMoney project on sourceforge with dedicated tools for online interactions (number 4). The module is now available. (number 5). This process corresponds also to the individual evolution of the user-champion who became a developer of the python programming language.

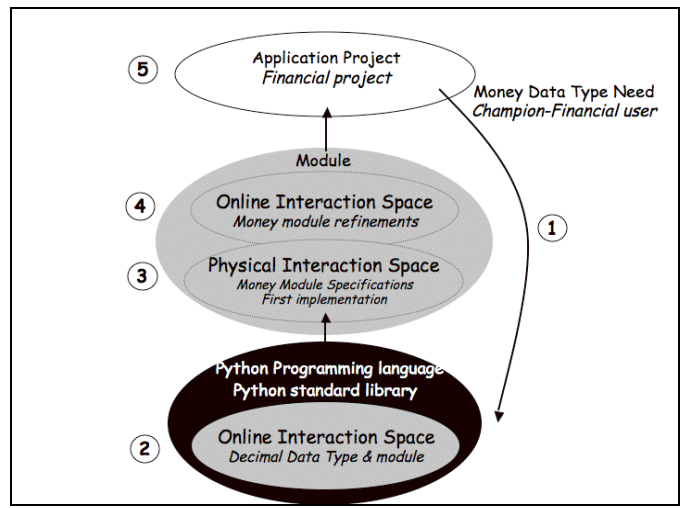

**Figure 3**. Broader view of the PEP 327 design process

## 5 Conclusion, Limits and Further works

In this work, we have outlined that the Python project is in fact the core of a galaxy of related application domain modules and projects, in which Python users are involved and that they may participate to the design process in both online and physical interactions spaces. The relative importance and the nature of activities occurring in these spaces are still to be characterized more precisely, such as the forms of participation of the users in these spaces. In particular, we can go closer in the understanding of the design and use of modules adapted to users needs.

A case study of a pushed-by-user proposal (PEP) highlighted that the process is distributed in the discussion space (in the python-list for users and the python-dev mailing-lists) and in the physical interactions spaces, with a specialisation of design topics in these spaces. But several issues still need to be explored. We can wonder what occurred before the first post by the user-champion in the Python Interactions spaces (with private e-mails or chat for instance). We have still to go on to precise the form of participations occurring during the phase of specification of the PEP in the online interaction spaces, specifying the participants involved and the nature of this involvement. We have to characterize the nature of the support the user-champion received from other participants. For instance, the user-champion told, publicly and during our interview, that he received technical and English (he is from Argentina) help from some specific administrators and developers.

In this direction, we have to characterize more finely the cross-participation process highlighted in this case study where some cross-participants have been identified. They may link users and developers communities (communities from application domain and the Python "core" community), acting as boundary spanners [24] between them. This implies to complete content analysis of messages occurring in parallel discussions.

Finally, to be able to generalise our research results, this study has to be extended to other pushed-by-users PEP in the Python project, and in other OSS projects.

## Acknowledgments

The authors wish to thank the reviewers for their helpful comments and suggestions.## References

1. Ducheneaut, N. (2005). Socialization in an Open Source Software Community: A Socio-Technical Analysis. *Journal of Computer Supported Collaborative Work, 14*, 323-368..
2. de Souza Sieckenius, C., and Preece, J. (2004). A framework for analysing and understanding online communities. *Interacting with computer, 16*, 579-610.
3. Détienne, F., Martin, G., and Lavigne, E. (2005). Viewpoints in co-design: A field study in concurrent engineering. *Design Studies, 26 (3),* 215–241.
4. Stempfle, J., and Badke-Schaub, P. (2002). Thinking in design teams - an analysis of team communication. *Design Studies, 23*, 473–496.
5. Gasser, L., Scacchi, W., Ripoche, G., and Penne, B. (2003). *Understanding Continuous Design in F/OSS project*. Communication at ICSSEA-03, Paris, France, December 2003.
6. Visser, W. (2004). *Dynamics aspects of design: Elements for a cognitive model of design*. Research report 5144, INRIA, Rocquencourt, France.
7. Carroll, J-M. (1996) Encountering others: reciprocal openings in participatory design and user-centered design. *Human-Computer Interaction, 11*, 285-290.
8. Scacchi, W. (2001). Understanding the requirements for developing Open Source Software Systems. *IEE Proceedings--Software, 149(1)*, 24-39.
9. Mockus, A., Fielding, R. T., and Herbsleb, J. (2002). Two Case Studies of Open Source Software Development: Apache and Mozilla. ACM Transactions on Software Engineering and Methodology, 11(3), 309–346.